\documentstyle[prl,aps,multicol,epsfig,amssymb]{revtex}
\newcommand{\putFig}[3]{\begin{figure}[t]%
\centering\epsfig{file=#1,width=0.983\linewidth}\nopagebreak\\[1ex]%
\begin{minipage}{0.994\linewidth}\caption{#3}\end{minipage}%
\label{#2}\end{figure}}
\newcommand{\figurA}{%
\putFig{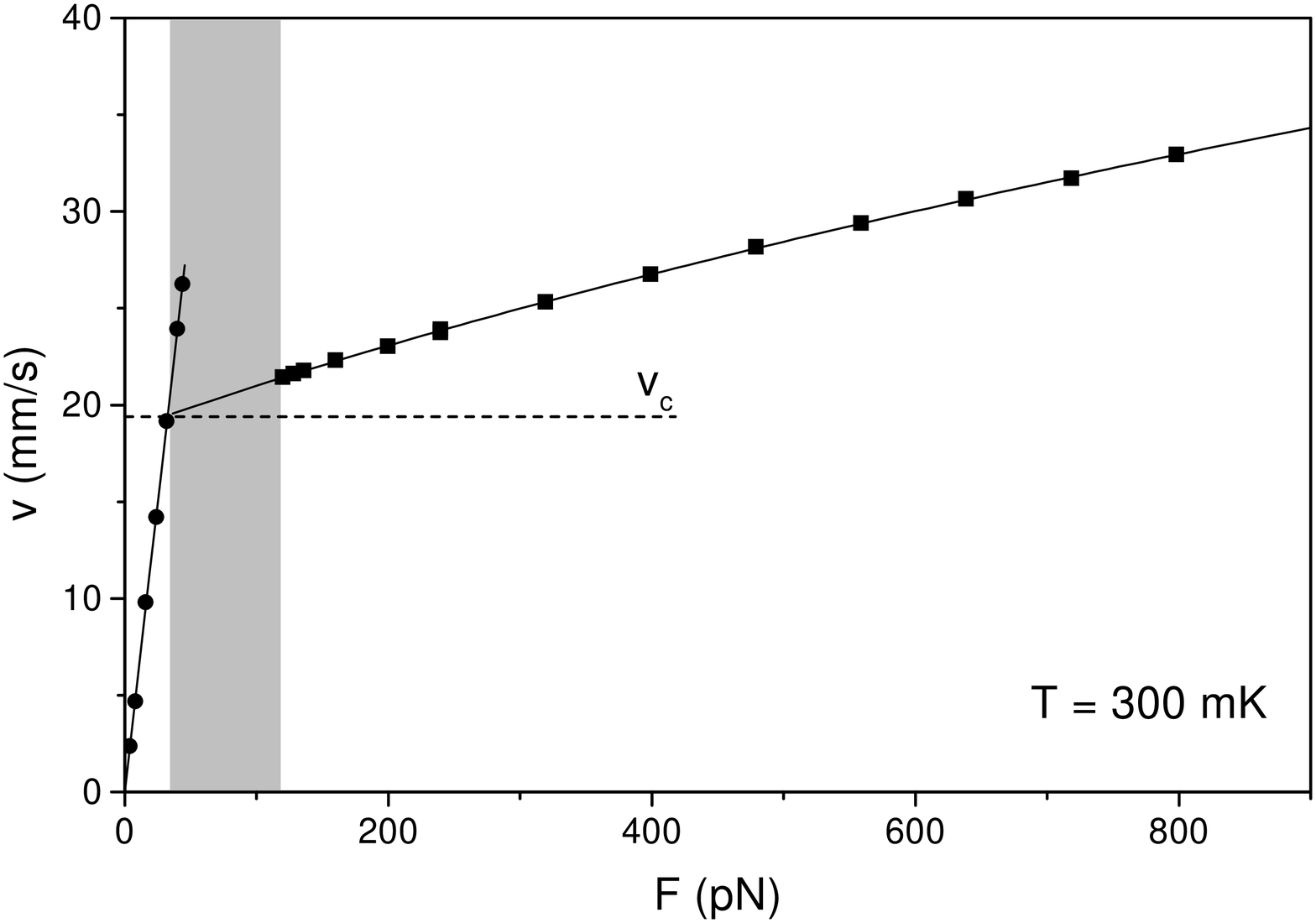}{Figure1}{%
Velocity amplitude of the oscillating sphere as a function of the driving 
force at $300\,\mbox{mK}$. At small drives the flow is laminar and the linear 
behavior is determined by ballistic phonon scattering (\textbullet). The 
shaded area marks the regime of instability above the critical velocity 
$v_c$. At larger driving forces we observe stable nonlinear turbulent drag 
({\tiny $\blacksquare$}), where the solid line is a fit of the quadratic 
drag force $F_D$, 
see text.}}

\newcommand{\figurB}{%
\putFig{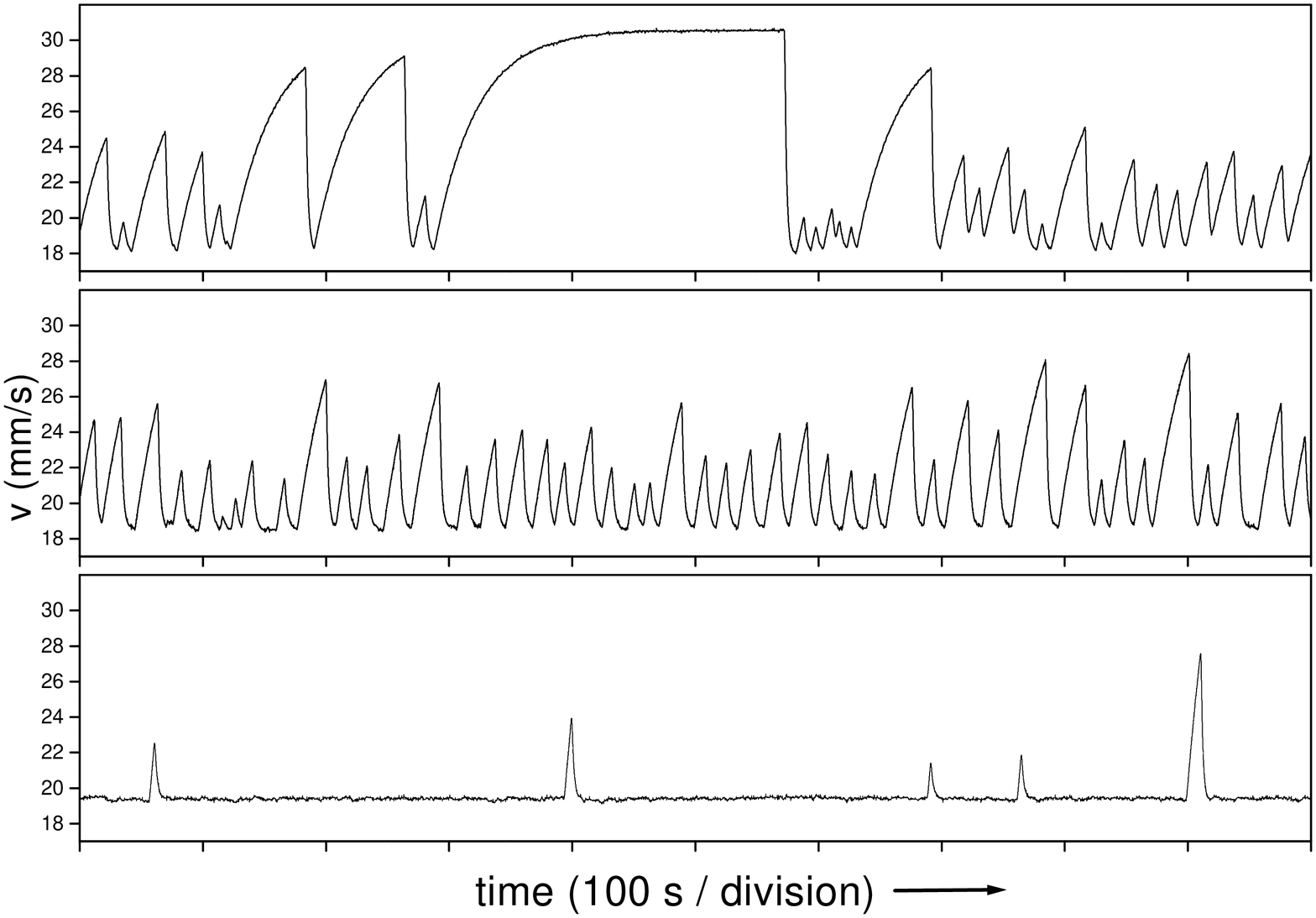}{Figure2}{%
Intermittent switching of the velocity amplitude at $300\,\mbox{mK}$. The low 
level corresponds to turbulent flow while the increase occurs during a 
laminar phase. The driving forces (in \mbox{pN}) are 47, 55, 75 (from top to 
bottom).}}

\newcommand{\figurC}{%
\putFig{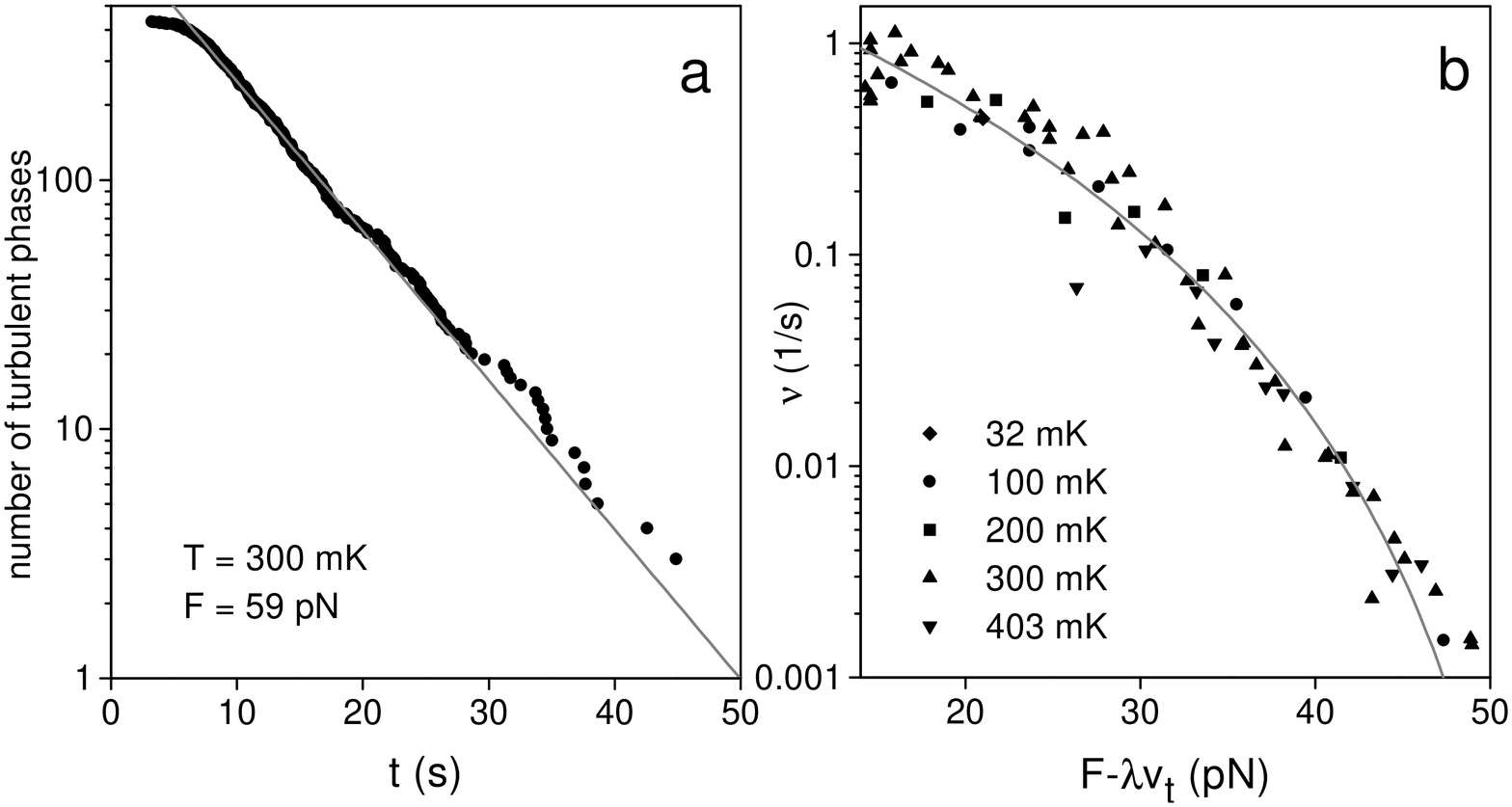}{Figure3}{%
Statistical analysis of the lifetimes $t$ of the turbulent phases. a) The 
exponential distribution $P(t)$ at a given temperature and driving force. b) 
The slopes $\nu$ measured at different temperatures as a function of the 
turbulent drag force. The solid line is a fit of a fourth power divergence 
of $1/\nu$ at a critical force of $54.6\,\mbox{pN}$.}}

\newcommand{\figurD}{%
\putFig{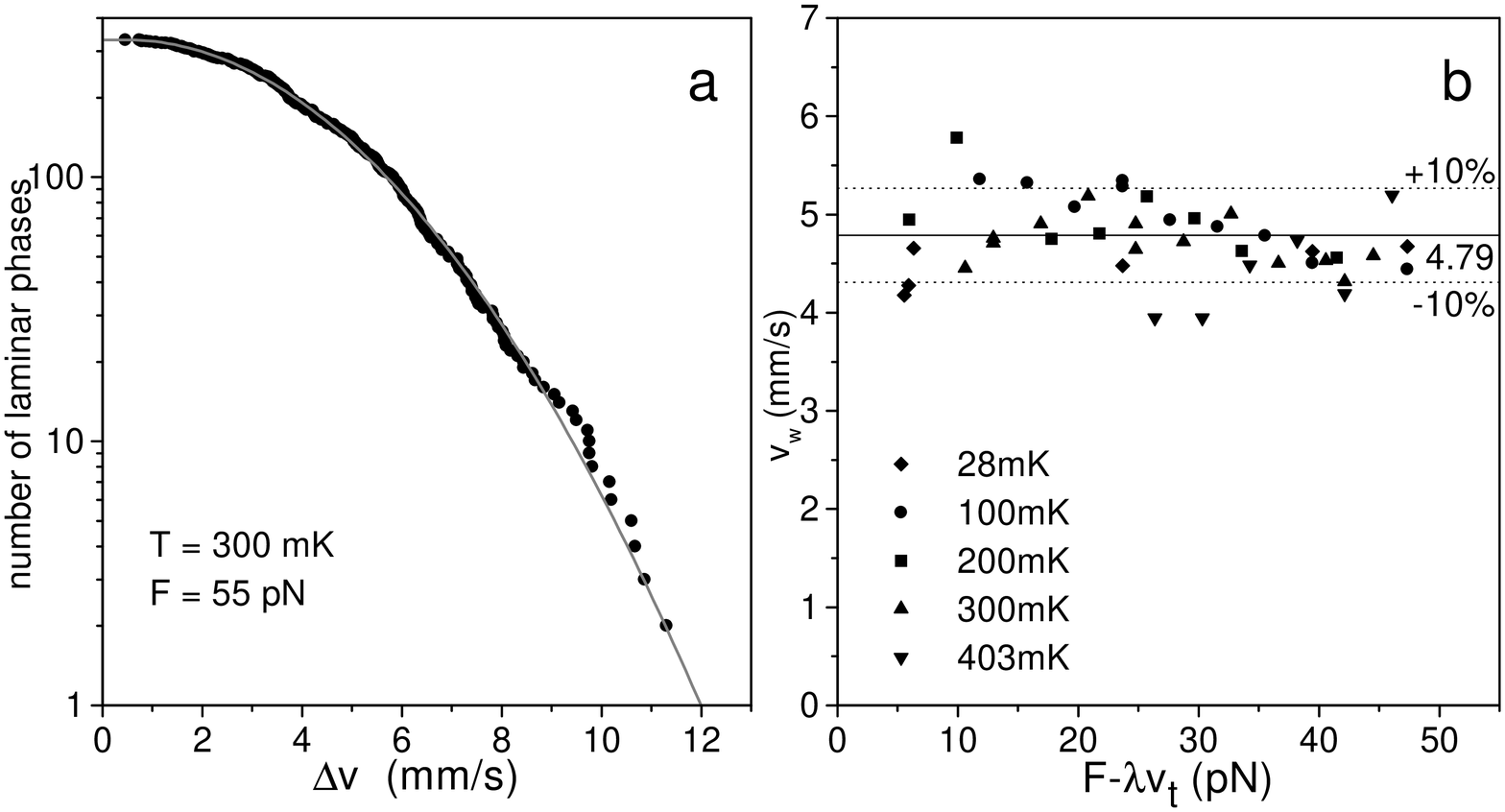}{Figure4}{%
Statistical analysis of the laminar phases. a) The Gaussian distribution 
$P(\Delta v)\propto \exp(-(\Delta v / v_W)^2)$ 
of the velocity increase $\Delta v$ at a given temperature and driving 
force. b) The fitting parameter $v_W$ is shown to be independent of 
temperature and driving force.}}

\newcommand{\figurE}{%
\putFig{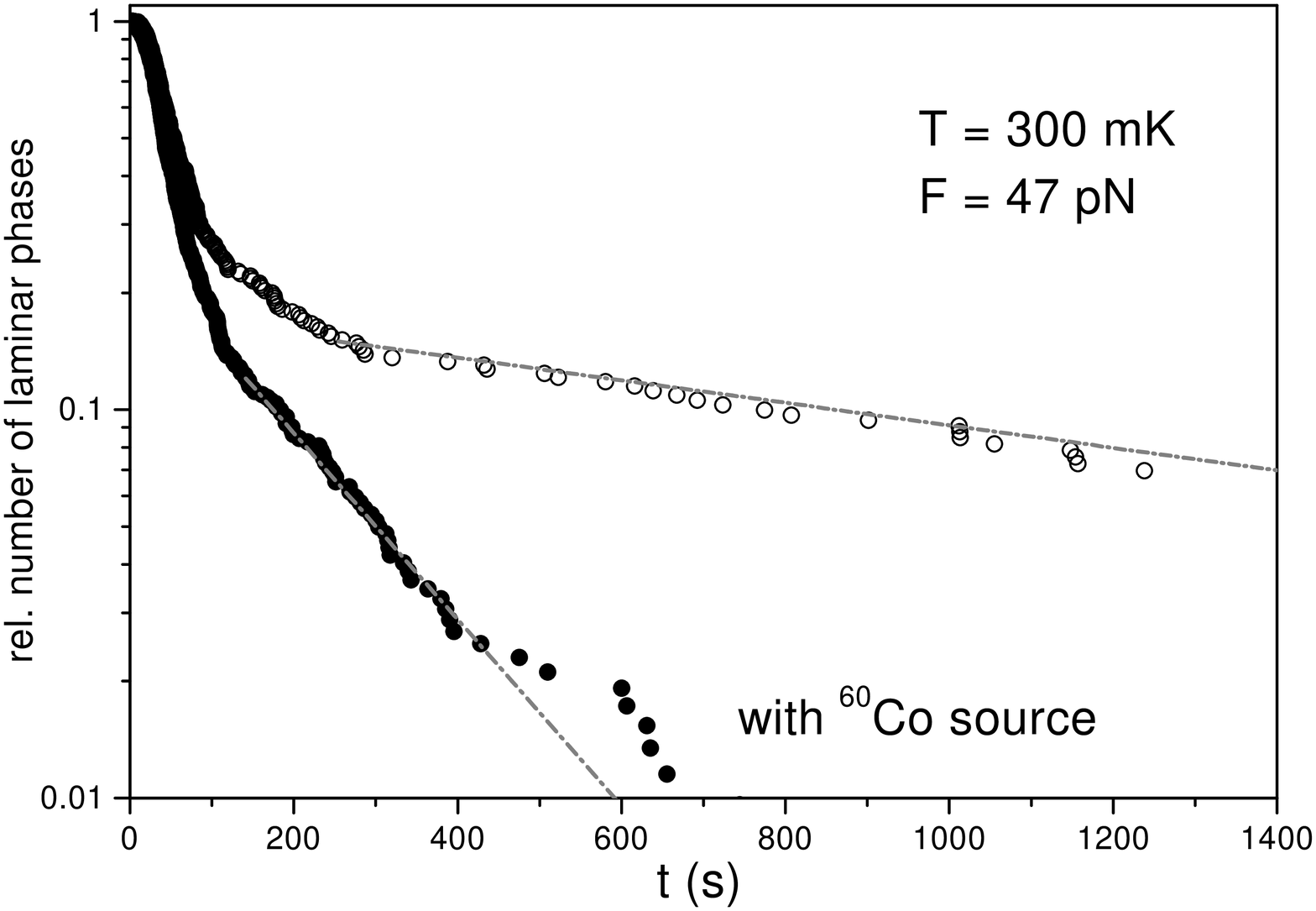}{Figure5}{%
Distribution of the lifetimes of the laminar phases slightly above the 
critical velocity $v_c$. For $t \gg \tau = 31\,\mbox{s}$ (at 300\,mK) the 
mean lifetime of 25 minutes is reduced by the radioactive source to 3 
minutes.}}
\begin {document} 
\draft
\title{
Intermittent Switching between Potential Flow and Turbulence in Superfluid 
Helium at mK Temperatures
}
\author{Michael Niemetz, Hubert Kerscher, and Wilfried Schoepe}
\address{Institut f\"ur Experimentelle und Angewandte Physik,
Universit\"at Regensburg, D-93040 Regensburg, Germany}
\date{\today}
\maketitle
\begin {abstract}
Superfluid flow around an oscillating microsphere is investigated at 
temperatures down to 25\,mK. Stable laminar flow below a critical velocity 
and turbulence at large drives are found to be separated below 0.5\,K by an 
intermediate range of driving forces where the flow is unstable, 
intermittently switching between laminar and turbulent phases. We have 
recorded time series of this switching phenomenon and have made a 
statistical analysis of the switching probability. The mean lifetime of the 
turbulent phases grows with increasing drive and becomes infinite at a 
critical value. Stability of the laminar phases  above the critical velocity 
is limited by natural background radioactivity or cosmic rays.
\end{abstract}


\begin{multicols*}{2}  
Turbulence in a classical liquid is based on the nonlinearity of the 
Navier-Stokes equation of motion \cite{frisch} while turbulence in 
superfluid helium, which also has been a very active area of research for 
several decades, is described by two-fluid hydrodynamics of the normal and 
superfluid components of the liquid \cite{donnelly,vts}. Because of the quantum mechanical properties of the liquid, turbulence in 
superfluid helium consists of a tangle of individual vortex lines having 
the same quantized circulation $h/m_4$, where $h$ is Planck's constant and 
$m_4$ is the mass of a helium-4 atom. In this respect, superfluid 
turbulence is different from
classical turbulence but it has also striking similarities like a Kolmogorov 
energy spectrum in the inertial range \cite{stalp}. Both cases represent 
different branches of the general phenomenon of fluid turbulence. 
A completely new aspect arises when 
superfluid turbulence is considered in the limit of very low temperatures 
\cite{ssv}. In this case all what is left of the normal fluid component is 
a very dilute gas of ballistically propagating phonons and therefore 
turbulence in the pure superfluid can be investigated. While a number of 
experiments have been performed which demonstrate the production of 
individual vortex lines both in superfluid $^4$He and in 
superfluid $^3$He \cite{packard} and the decay of the vortex tangle 
\cite{mcclintock}, the transition to turbulence in the flow of the 
superfluid around a macroscopic body so far has not been investigated at 
these very low temperatures.

In our present work, we have studied superfluid flow around an oscillating 
microsphere at temperatures down to 25\,mK. Laminar and turbulent flow can 
be identified by the drag force on the sphere, which in case of laminar flow 
is linear and determined by ballistic phonon scattering, while for 
turbulent flow, which exists above a critical velocity, the drag force is 
nonlinear and much larger \cite{jj1}. Here, we report on the observation of 
an intermediate regime located between stable laminar flow below the 
critical velocity and stable turbulence above a critical driving force, 
where the flow is found to be unstable, switching intermittently between 
laminar and turbulent phases. We have measured time series of this switching 
phenomenon at different constant temperatures and driving forces. We find, 
firstly, the mean lifetime of the turbulent phases to grow with increasing 
drive and to diverge at a critical value. Secondly, the stability of the 
laminar phases above the critical velocity is analyzed and the probability 
for switching to turbulence is obtained. Surprisingly, we find the 
long term stability of the laminar phases close to the critical 
velocity to be limited ultimately by natural radioactivity or cosmic rays.

Our experimental technique has been described in earlier work \cite{jj1} and 
is only briefly summarized here as follows. The 
microsphere (radius $R = 0.1\,\mbox{mm}$, mass $m = 27\,\mu\mbox{g}$) 
is made of ferromagnetic 
$\mathrm{SmCo_5}$. Therefore, we can use superconducting levitation, i.e., 
the repulsion of a permanent magnet from a superconducting surface, to let 
it float 
in the middle between the superconducting electrodes of a parallel 
plate capacitor made of niobium (spacing $d=1\,\mbox{mm}$, diameter $4\,\mbox{mm}$) 
without any mechanical suspension. Lateral 
stability is provided by flux lines trapped in the superconductor. 
The sphere carries an electric charge $q$ of about 1\,pC when it 
levitates, because we apply a voltage of several hundred volts to the bottom 
electrode when cooling the capacitor through the critical temperature 
$T_c=9.2\,\mbox{K}$ of niobium. 
Vertical oscillations of the sphere about its equilibrium 
position can now be excited by applying a small ac voltage $U_{ac}$ ranging 
from $0.1\,\mbox{mV}$ up to $20\,\mbox{V}$ at the resonance frequency of the sphere 
$\sim 120\,\mbox{Hz}$. These oscillations induce an ac current $q v / d$ in the 
electrodes, where $v$ is the velocity amplitude, which is detected by an 
electrometer. The damping of the oscillations is very small, $Q$-values in 
excess of one million are achieved if the capacitor is evacuated 
\cite{gloos}. We then fill the capacitor with pure superfluid 
helium\cite{pure} and measure the velocity amplitude at resonance as a 
function of the driving force $F = q U_{ac} /d$ at constant temperature 
between 25\,\mbox{mK}{} and 1\,\mbox{K}. 

As reported earlier \cite{jj1} there is a linear regime $v(F)$ at 
small amplitudes, see Fig.\,1. This indicates a linear drag force 
$\lambda (T) v$ where the drag coefficient $\lambda \propto T^4$ is 
determined by ballistic scattering of thermal phonons which can be 
considered as a dilute gas of excitations whose density rapidly 
decreases as $T^4$. The flow of the superfluid around the sphere is 
frictionless potential flow. At large driving forces, however, the sphere 
creates vorticity in the superfluid. 
Now the drag force is large and nonlinear, see Fig.\,1. We find the drag 
force to be given by $F_D = \gamma v^2 - F_0$, where $\gamma = c_D \rho \pi 
R^2 / 2$ is the classical turbulent drag coefficient ($\rho$ is the density 
of the liquid and $c_D \approx 0.4$ for a sphere) and $F_0$ is a constant. 
In contrast to a classical liquid where $F_0=0$ in our case turbulent drag 
($F_D>0$) exists only above a critical velocity $v_c = 19.4\,\mbox{mm/s}$.
A similar drag force has recently been calculated for a cylinder in a 
two-dimensional dilute Bose-Einstein condensate, although the critical 
velocity is very much different in that case \cite{adams}.
\figurA

Most interesting is our new observation of an intermediate regime where 
neither potential flow nor turbulence are stable but instead the flow 
switches between both patterns intermittently \cite{lt22}. This regime of 
instability is observed only below $0.5\,\mbox{K}$ and
extends from the critical velocity $v_c$ up to a critical driving force 
where enough power is being delivered to the superfluid turbulence 
to become stable. We have recorded time series (up to 36 hours per run) 
of this switching phenomenon at various driving forces and temperatures. An 
example is shown in Fig.\,2 where a section of 17 minutes is 
shown for three different driving forces at $300\,\mbox{mK}$. The amplitude 
switches between a low level corresponding to turbulent drag and an 
exponential recovery of the laminar level corresponding to phonon 
scattering. If the laminar phase lasts long enough the stationary value for 
laminar flow is reached (see upper time series). When the laminar flow 
breaks down a rapid drop to the turbulent level occurs. It is obvious from 
Fig.\,2 that with increasing driving force the lifetimes of the 
laminar phases shrink while those of the turbulent phases grow. 

\figurB
\figurC

In the following we discuss the results of a statistical analysis of these 
time series based on reliability theory \cite{gnedenko}.
A typical distribution of the lifetimes of the turbulent phases is shown in 
Fig.\,3a where the number $P$ of turbulent phases having a lifetime longer 
than a given threshold $t$ is plotted on a logarithmic scale versus $t$. We 
always find an exponential distribution $P(t) \propto \exp(-\nu t)$ with 
$1/\nu$ being the mean lifetime. This implies a probability for the 
breakdown of the turbulent phase ("failure rate") $- d\ln P / dt = \nu$ 
being constant in time (i.e no history dependence). The values of $\nu$ are 
independent of temperature and collapse to an universal drive dependence if 
the strongly temperature dependent laminar drag $\lambda v$ (where $v=v_t$ 
is the velocity amplitude of the turbulent phase) is subtracted from the 
external driving force. Obviously, $\nu$ depends only on the turbulent drag 
force $F-\lambda v_t$ and vanishes at a critical drive 
approximately as the fourth power of the distance from a critical value,
see Fig.\,3b. At the critical drive the power delivered to the 
turbulent superfluid is $0.6\,\mbox{pW}$ which corresponds approximately to the 
production of one vortex ring of the size of the sphere per half period of 
the oscillation. 

\figurD

Stability of the laminar phases can be analyzed in two ways, either the 
distribution of the signal height $\Delta v$ (velocity at breakdown minus 
turbulent velocity $v_t$) is considered or the distribution of the 
lifetimes. Both variables are related by the exponential recovery 
$\Delta v (t) = \Delta v_{max} (1-\exp(-t /\tau))$, where $\Delta v_{max}$ 
is the difference between the stationary laminar velocity amplitude 
$F/\lambda$ and $v_t$, and $\tau = 2m/\lambda$ is the time 
constant. In Fig.\,4a the distribution of the signal height $\Delta v$ is 
shown to be of the Gaussian form $P(\Delta v) \propto \exp(-(\Delta v / 
v_W)^2)$ where $v_W$ is a fitting parameter. This implies a failure rate
(per unit velocity increment) $-d\ln P/d\Delta v= 2 \Delta v / v_W^2$ to be 
proportional to the signal height $\Delta v$. The corresponding 
probability density of $\Delta v$ ($-d P / d\Delta v$) is called a Weibull 
distribution. The fitting parameter $v_W \approx 4.8\,\mbox{mm/s}$ is 
found to be completely independent of both temperature and driving force, 
see Fig.\,4b. We expect it to depend on the size or the surface 
roughness of the sphere but experiments with different spheres remain to be 
performed in the future. 
Because $\Delta v(t)$ saturates for $t \gg\tau$, it is more informative in 
this case to analyze the lifetime distribution, in particular near $v_c$ 
where the laminar phase becomes stable. From $P(t) = P(\Delta v(t) ) \propto 
\exp(-(\Delta v(t) / v_W)^2)$ we conclude that for $t \gg\tau$ $P(t)$ must 
be constant and therefore the failure rate (per unit time) is $-d\ln P / dt 
= v_W^{-2} d(\Delta v)^2 / dt = 0$. This simply means that if the laminar 
phase has survived for a time $t\gg\tau$ it will be stable indefinitely 
although the velocity is clearly above $v_c$. Our experimental results, 
however, demonstrate that the lifetime is finite: $\ln P(t)$ slowly 
decreases at large $t$ and from a straight line fit to the data in Fig.\,5 
we obtain a mean lifetime of 25\,minutes. We can rule out mechanical 
vibrations of the cryostat as the source of this instability: jumping on the 
floor, slamming the door, even transferring liquid helium to the cryostat 
had no effect. Therefore, another mechanism must be responsible. In fact, 
placing a small radioactive source $^{60}\mathrm{Co}$ ($74\,\mbox{kBq}$) near the 
cryostat has a dramatic effect: the mean lifetime of the laminar phase is 
reduced to 3.0 minutes. We have measured the dose rate of the source at the 
position of the measuring cell inside the cryostat (taking into account a 
measured 20\% loss in the dewar walls) to be 440\,nGy/h ($\pm$\,5\%). 
Comparing this value with a measured dose rate due to natural background 
radiation in our laboratory of 50\,nGy/h ($\pm$\,10\%), which is typical for 
our area, we obtain an increase of the dose rate due to the source by a 
factor of $(440+50)/50=9.8$. Within the error bars this compares well with 
the measured reduction of the mean lifetime of the laminar phases by a 
factor of $25/3.0=8.3$.
This effect may be attributed to the creation of ions in the superfluid, 
which produce local vorticity either during the creation and recombination 
processes or when the ions are accelerated by the electric field which 
exists in our measuring cell \cite{rr}. This vorticity then initiates the 
breakdown of the potential flow if the velocity amplitude of the sphere is 
larger than the critical velocity $v_c$. We therefore conclude that the 
lifetime of the metastable laminar phases which we observe above $v_c$ is 
only limited by natural background radioactivity. 

\figurE

In summary, we have found that the transition from potential flow to 
turbulence around a sphere in superfluid helium below 0.5\,K occurs by 
intermittent switching between both flow patterns. We understand now why above 0.5\,K the intermittent switching changes into 
the hysteretic behavior observed earlier \cite{jj1}: the velocity 
increases $\Delta v$ become very small at higher phonon drag which implies 
a very low failure rate of the laminar phase. But if it fails (i.e. when 
the driving force $F$ is largely increased) the following turbulent phase 
is stable because the critical drive is exceeded. 

At present we have no theoretical model which could describe our experimental results for the 
failure rates of the laminar and turbulent phases. But because the 
transition to turbulence in superfluid helium at mK temperatures is a new 
phenomenon, we hope that our results will stimulate further theoretical 
progress.

We are most grateful to H. von Philipsborn for measuring the dose rates and 
for illuminating discussions. We thank M. Creuzburg for clarifying comments 
on dosimetry and J. Reisinger for the loan of the radioactive source. 
We had helpful discussions on turbulence with H.R. Brand. This work is 
supported by the Deutsche Forschungsgemeinschaft.

\end{multicols*} 


\begin{thebibliography}{99}
\bibitem{pure} We used isotopically pure helium-4 (impurity concentration 
$<$\,1\,ppb) purchased from Campro Scientific GmbH, Berlin.
\end{thebibliography}
\end{document}